 \documentclass[switch]{article} 
 
 \usepackage{preprint}
 
 \usepackage{amsmath, amsthm, amssymb, amsfonts}
 
 \usepackage[numbers,square]{natbib}
 \usepackage[utf8]{inputenc}	
 \usepackage[T1]{fontenc}	
 \usepackage{xcolor}		
 \usepackage[hidelinks]{hyperref}	
 \usepackage{booktabs} 		
 \usepackage{nicefrac}		
 \usepackage{microtype}		
 \usepackage{lineno}		
 \usepackage{float}			
 \usepackage{graphicx}
 \usepackage{lipsum}		
 \usepackage[bbgreekl]{mathbbol}
 \usepackage{tabularx}
 \usepackage{subcaption}
 \usepackage{mathtools}
 \DeclarePairedDelimiter\abs{\lvert}{\rvert}%
 \DeclarePairedDelimiter\norm{\lVert}{\rVert}%
 \makeatletter
 \let\oldabs\abs
 \def\abs{\@ifstar{\oldabs}{\oldabs*}}
 \let\oldnorm\norm
 \def\norm{\@ifstar{\oldnorm}{\oldnorm*}}
 \makeatother

 \usepackage[ruled,resetcount]{algorithm2e}
 \makeatletter
 \renewcommand*\l@algocf{\l@figure}
 \makeatother
 \SetAlgoInsideSkip{smallskip}
 \SetAlCapNameFnt{\footnotesize}
 \SetAlCapFnt{\footnotesize}
 
 \graphicspath{{Figs/}}
 
 \usepackage{newfloat}
 \DeclareFloatingEnvironment[name={Supplementary Figure}]{suppfigure}
 \usepackage{sidecap}
 \sidecaptionvpos{figure}{c}
 
 \usepackage{titlesec}
 \titlespacing\section{0pt}{12pt plus 3pt minus 3pt}{1pt plus 1pt minus 1pt}
 \titlespacing\subsection{0pt}{10pt plus 3pt minus 3pt}{1pt plus 1pt minus 1pt}
 \titlespacing\subsubsection{0pt}{8pt plus 3pt minus 3pt}{1pt plus 1pt minus 1pt}
 
 \title{Improving pixel differentiation in holographic images}
 
 \usepackage{authblk}
 
 \author[1,*]{Peter J. Christopher}
 \author[1]{Ralf Mouthaan}
 \author[2]{John P. Freeman}
 \author[1]{Timothy D. Wilkinson}
 \affil[1]{Centre of Molecular Materials, Photonics and Electronics, University of Cambridge, UK}
 \affil[2]{BAE Systems}
 \affil[*]{pjc209@cam.ac.uk}

\begin{document}

     \maketitle

     \begin{abstract}
         Computer generated holography (CGH) has seen a resurgence in recent years due, in part, to the rise of virtual and mixed reality systems. The majority of approaches for CGH are based on a sampled Discrete Fourier Transform (DFT) and ignore the interstitial behaviour between sampling points in the replay field. In this paper we demonstrate that neighbouring replay field pixels can interfere significantly giving the visual impression of pixel movement and increased noise. We also demonstrate that increasing the separation between target pixels reduces the interference and improves pixel quality. This phenomena is demonstrated experimentally with close agreement between model and measured result.
         
         This work begins by introducing the concept of pixel differentiation before showing simulated models of pixel differentiation issues. Two mitigation approaches are introduced and an experimental system is then used to validate the simulation. Finally results are discussed and conclusions drawn.
     \end{abstract}
     \vspace{0.35cm}

     \normalsize
     \twocolumn
            
    \section{Introduction}
        
    Since its conception in the 1980s computer generated holography (CGH) has seen use in applications including lithography~\cite{Turberfield2000, Purvis2014}, displays~\cite{Maimone2017,Yamada2018}, imaging~\cite{Svoboda2013, Frauel2006, Sheen2001} and optical manipulation~\cite{Grier06,Melville2003,Grieve2009}. CGH is enabled by the use of spatial light modulators (SLMs) that modulate either the phase or amplitude of a coherent light source in order to shape the resulting diffraction pattern. The practical constraints of SLM modulation is the primary constraint in finding suitable diffraction patterns corresponding to a target replay field.
    
    A wide array of algorithms exist for adapting an ideal diffraction pattern to one achievable on real-world devices including simulated annealing \cite{kirkpatrick1983optimization}, direct binary search \cite{DirectSearch_2} and Gerchberg-Saxton \cite{gerchberg1972practical}. Real-world issues such as sampling, speckle and noise mean that expertise is required to use these techniques. 
    
    In this paper we introduce the issue of pixel differentiation. We start by presenting a brief background to CGH before presenting examples of pixel differentiation issues. We continue to discuss the impact of different performance factors and to present potential solution methods. Finally we discuss our findings and draw conclusions.

    \section{Background}
    
    Traditional approaches to CGH are based on the use of the Discrete Fourier Transform (DFT),
    
    \begin{align}
    & F_{u,v} = \mathcal{F}\{f_{x,y}\}         = \\
    & \frac{1}{\sqrt{N_xN_y}}\sum_{x=0}^{N_x-1}\sum_{y=0}^{N_y-1} f_{xy}e^{-2\pi i \left(\frac{u x}{N_x} + \frac{v y}{N_y}\right)} \label{fouriertrans2d5c}   \nonumber\\
    & f_{x,y} = \mathcal{F}^{ - 1 }\{F_{u,v}\} ] = \\
    & \frac{1}{\sqrt{N_xN_y}}\sum_{u=0}^{N_x-1}\sum_{v=0}^{N_y-1} F_{uv}e^{2\pi i \left(\frac{u x}{N_x} + \frac{v y}{N_y}\right)}  \label{fouriertrans2d5d} \nonumber
    \end{align}
    
    where $u$ and $v$ represent the spatial frequencies and $x$ and $y$ represent the source coordinates. For reasons of performance, DFTs for holography are typically calculated using Fast Fourier Transforms (FFTs) calculation times of $O(N_xN_y\log{N_xN_y})$ where $N_x$ and $N_y$ are the respective $x$ and $y$ resolutions~\cite{carpenter2010graphics,frigo2005design}.
    
\begin{figure}[tb]
        \centering
        {\includegraphics[trim={0 0 0 0},width=\linewidth,page=1]{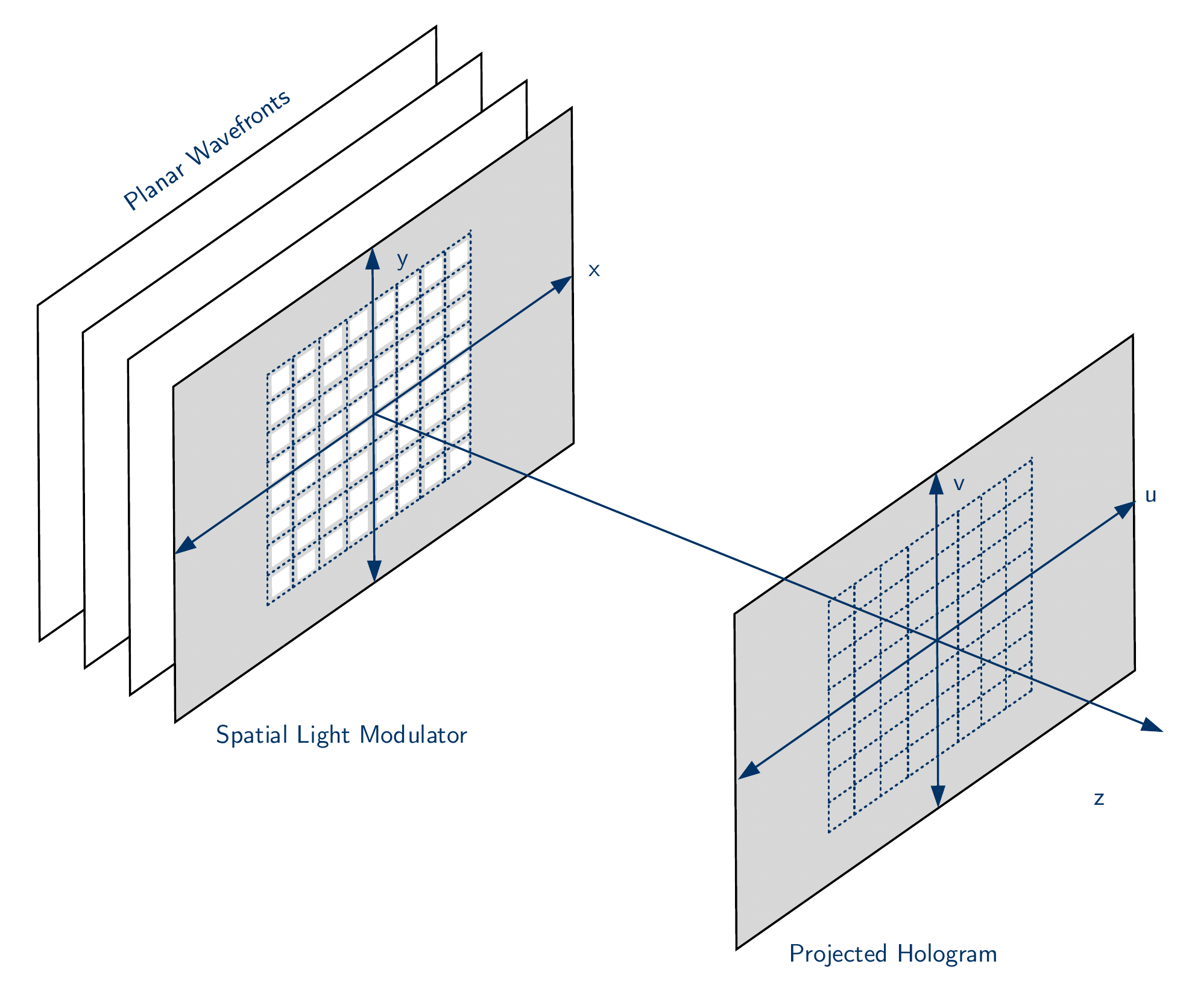}}
        \caption{Hologram coordinate systems. Used with permission from: \cite{MostChangedPixel}}
        \label{fig:figure01}
    \end{figure}

    If the SLM aperture is modelled as an array of point impulse functions, the resultant far-field replay pattern caused can be treated as the DFT of the SLM  pointwise multiplied by the illumination and pixel shape functions~\cite{goodman2005introduction}. The coordinate systems used are shown in Figure~\ref{fig:figure01} and we often refer to the projected image as the \textit{Replay Field} and the SLM as the \textit{Diffraction Field}.
        
    Real-world SLMs are capable of only limited modulation, typically amplitude-only or phase-only~\cite{Huang2018,de1997complex}. The challenge of generating an appropriate hologram then becomes that of finding function $f(x,y)$ conforming to the SLM constraints and to the relationship $F(u,v) = \mathcal{F}\{f(x,y)\}$ where $\mathcal{F}$ represents the Fourier transform.
    
    Additionally, it is worth noting that the target replay field values generated by the DFT are regularly sampled and the total intensity interstitially is not determined by the transform.
    
    \section{Pixel Differentiation}
    
\begin{figure}
        \centering
        {\includegraphics[trim={0 0 0 0},width=0.8\linewidth,page=1]{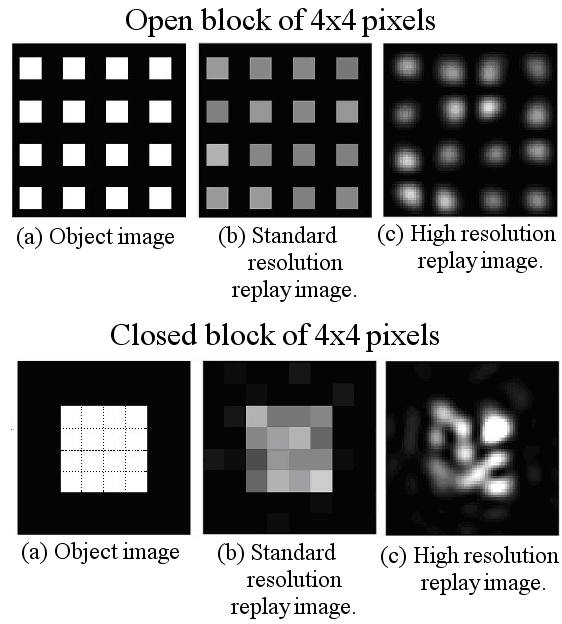}}
        \caption{Pixel differentiation for a continuous phase $128 \times 128$ pixel hologram. Only the central $8\times 8$ region of the image is shown.}
        \label{fig:figure02}
    \end{figure}

    To introduce the problem of pixel differentiation, two separate images are considered shown in Figure~\ref{fig:figure02}. 
    
    Figure~\ref{fig:figure02} (top, a) shows a grid of phase insensitive target pixels with a two pixel separation where the separation between pixels is taken as being the distance centre to centre. The expected replay field (b) using a DFT is shown along with a high resolution model (c). It can be seen that the shape of the replay field pixels poorly correspond to the DFT but are distinct. 
    
    The first image in Figure~\ref{fig:figure02} (bottom) shows an otherwise identical case (a) where the pixels have a one pixel separation. Here the expected DFT generated replay field (b) is similar to the previous but the individual pixels are now poorly differentiated. 
    
    In both cases the hologram portions shown form the central $8\times 8$ pixel region of a $128 \times 128$ image. The SLM pattern is taken as the superposition of gratings that form the individual pixels. This is modulated to the constraints of a phase SLM. The SLM pixels are here treated as point impulse functions.
    
    \section{The two-pixel case} \label{sec:two}
    
    To better understand the behaviour here we investigate the simplest possible case, two pixels displayed by a full continuous phase SLM. The phase of the first target pixel $\phi_1$ is kept constant while the phase of the second $\phi_2$ is varied in the interval $\left(0,2 \pi\right]$. It is assumed that the SLM is illuminated by a perfectly collimated, top-hat profile beam. 
    
    \subsection{Simultaneous}
    
\begin{figure}
        \centering
        {\includegraphics[trim={0 0 0 0},width=0.8\linewidth,page=1]{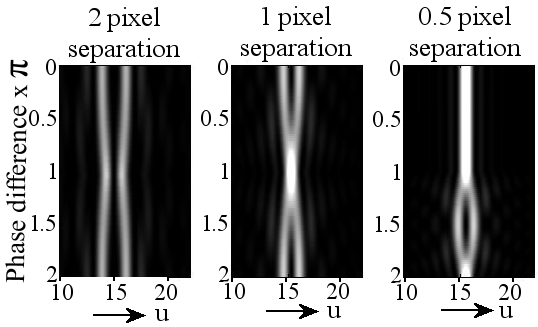}}
        \caption{Section through neighbouring pixels on a $128 \times 128$ pixel hologram against difference in target phase angle between the two pixels.}
        \label{fig:figure03}
    \end{figure}

\begin{figure}
        \centering
        {\includegraphics[trim={0 0 0 0},width=0.8\linewidth,page=1]{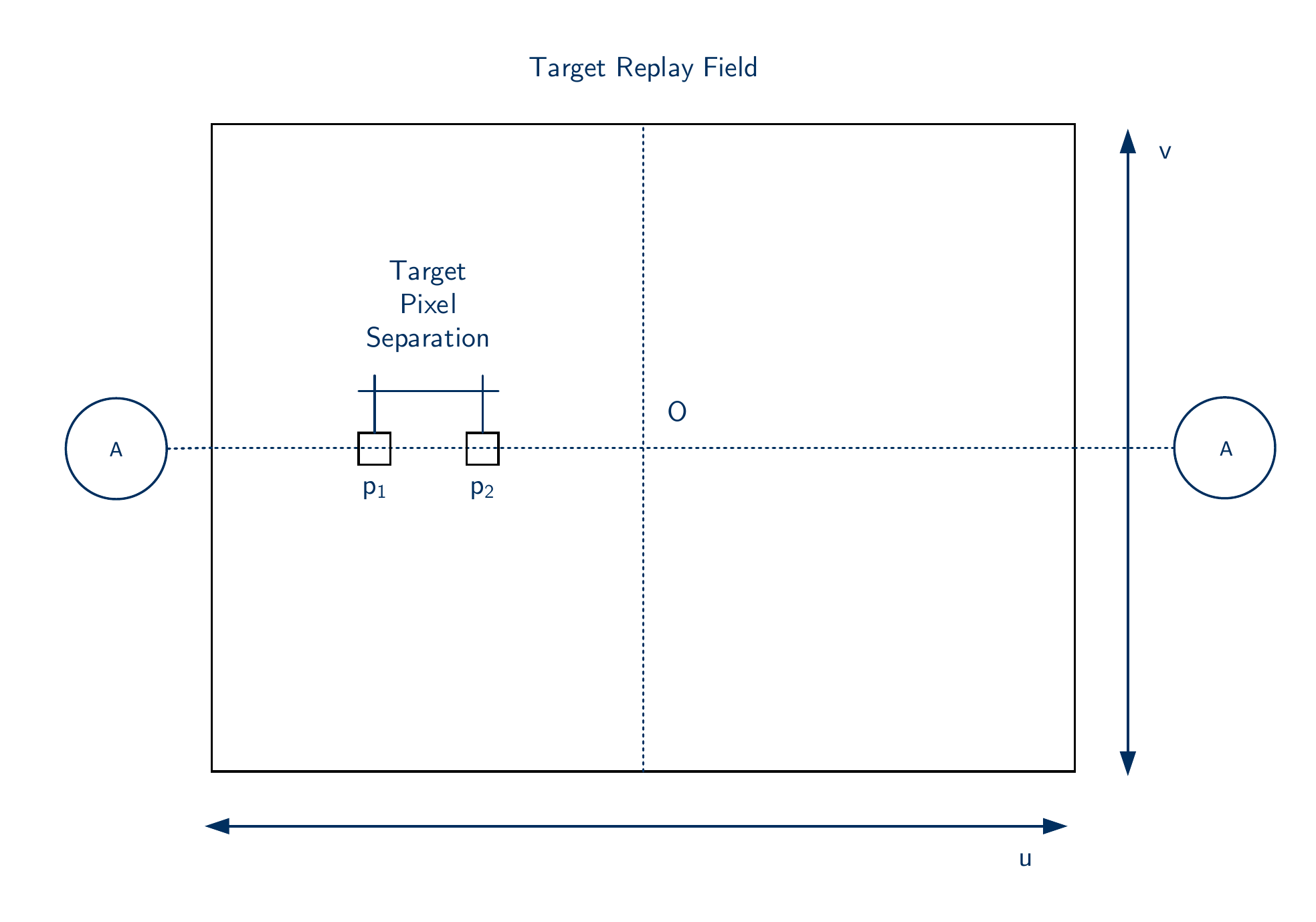}}
        \caption{Configuration of replay field cross section.}
        \label{fig:FreemanCoords}
    \end{figure}

    A cross section through the centres of the reconstruction against $(\phi_2-\phi_1)$ is shown in Figure~\ref{fig:figure03} for a configuration shown in Figure~\ref{fig:FreemanCoords} with cross section A-A. This shows the case of SLM pixels with a 2, 1, $\nicefrac{1}{2}$ pixel separation in (a), (b) and (c) respectively and shows that the two pixels can appear to move closer together or further apart depending on phase separation.
    
    The graph shows the distance along the $u$ axis in the $128 \times 128$ pixel hologram used for testing with $v$ being constant at zero. The SLM pattern is generated using superposition of constituent gratings and the cross-section is super-sampled. It will be noted that while the pixels are central in $v$ they are offset in $u$ for reasons discussed later.
    
    It is worth noting also that this effect is independent of the location of the pixel pair in the target image and is only influenced by their relative phase difference and their relative spacing.
            
    \subsection{Sequential}
    
\begin{figure}
        \centering
        {\includegraphics[trim={0 0 0 0},width=0.8\linewidth,page=1]{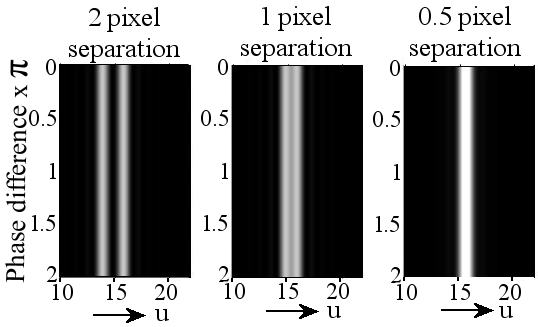}}
        \caption{Time averaged section through neighbouring pixels on a $128 \times 128$ pixel hologram against difference in target phase angle between the two pixels where the pixels are shown sequentially and the impulse is integrated.}
        \label{fig:figure04}
    \end{figure}

    For comparison, Figure~\ref{fig:figure04} shows the case of two pixels shown alternately and the result time averaged. Here the distinction is as expected and shows distinct pixels independently of phase without the apparent pixel movement. 
    
    \subsection{Interpretation}
    
    The difference between the simultaneous and sequential cases is stark and requires investigation. The value of a given location $H_{x,y}$ in the replay field is given as a function of target replay field values $T_{u,v}$
    
    \begin{equation}
    H_{x,y} = \sqrt{N_xN_y}T_{u,v} e^{\left[2\pi i\left(\frac{ux}{N_x}+\frac{vy}{N_y}\right)\right]}
    \end{equation}
    
\begin{figure}
        \centering
        {\includegraphics[trim={0 0 0 0},width=0.6\linewidth,page=1]{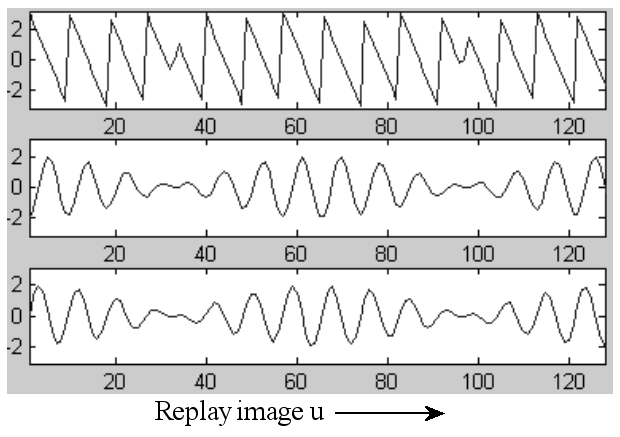}}
        \caption{1D Plot showing complex addition of 2 holograms of 2 pixels (real middle, complex bottom) resulting in the saw-tooth phase hologram with discontinuities (top).}
        \label{fig:figure05}
    \end{figure}

    It can be seen for a single target pixel that the required hologram is a phase grating with frequency determined by the location of the target pixel in the replay field. For two pixels, the two constituent gratings are summed. Here, however, behaviour similar to Figure~\ref{fig:figure05} is observed where the two frequencies constructively and destructively interfere in a \textit{beats}-like phenomenon. When modulated to the SLM constraints, this introduces discontinuities in the target phase profile.
    
\begin{figure}
        \centering
        {\includegraphics[trim={0 0 0 0},width=0.6\linewidth,page=1]{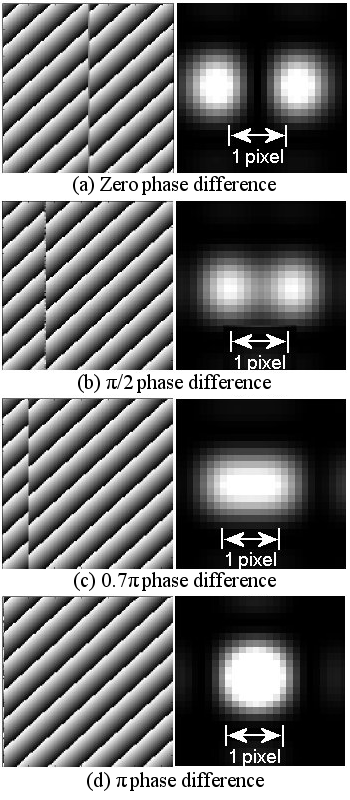}}
        \caption{$128 \times 128$ phase holograms (left) used to display the replay field portions (right) that target two neighbouring pixels with a single pixel separation and the given phase difference.}
        \label{fig:figure06}
    \end{figure}

    This is shown further in Figure~\ref{fig:figure06} where the phase holograms to generate two pixels are shown left against a close-up of the resultant image right. When the phase difference is set to $\pi$, the induced discontinuity is coincident with the hologram edge and manipulation of the phase difference between the gratings allows us to control the location of the discontinuity.
    
\begin{figure}
        \centering
        {\includegraphics[trim={0 0 0 0},width=0.6\linewidth,page=1]{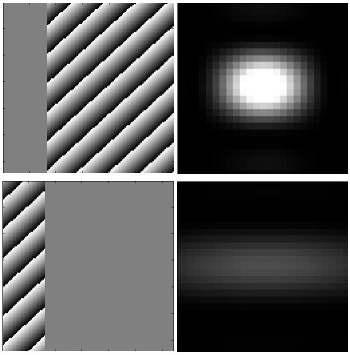}}
        \caption{Separation of the two portions of Figure~\ref{fig:figure06}b. The uniform grey portions are taken as being zero.}
        \label{fig:figure07}
    \end{figure}

    Examining Figure~\ref{fig:figure07} highlights the distinction between the two contributing components.
    
\begin{figure}
        \centering
        {\includegraphics[trim={0 0 0 0},width=0.8\linewidth,page=1]{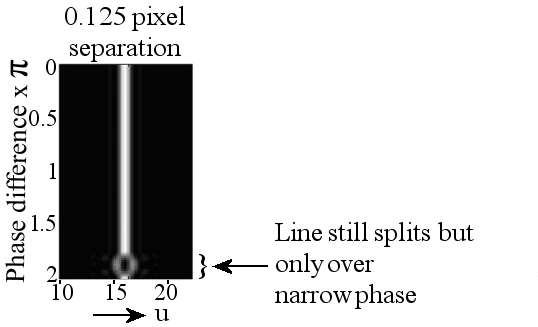}}
        \caption{Section through neighbouring pixels with a 0.125 pixel spacing on a $128 \times 128$ pixel hologram against difference in target phase angle between the two pixels.}
        \label{fig:figure08}
    \end{figure}

    If the distance between the target pixels is changed, the periodicity of the beats phenomena is changed but still exhibits the same pixel movement behaviour and discontinuities. For example Figure~\ref{fig:figure08} shows the identical case to Figure~\ref{fig:figure03} but with target pixels only separated by 0.125 pixel spacing. 
    
\begin{figure}
        \centering
        {\includegraphics[trim={0 0 0 0},width=0.8\linewidth,page=1]{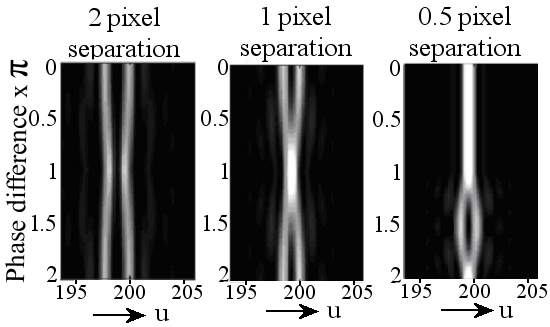}}
        \caption{Section through neighbouring pixels on a $128 \times 128$ pixel hologram against difference in target phase angle between the two pixels.}
        \label{fig:figure09}
    \end{figure}

    Additionally, this two pixel case is independent of the resolution of the SLM. For example, Figure~\ref{fig:figure03} is reproduced in Figure~\ref{fig:figure09} but for the case of a $1024\times 1024$ hologram.
    
    \section{Application}
    
    \subsection{Frame Averaging}
    
    The results in Section~\ref{sec:two} and in particular Figure~\ref{fig:figure04} suggest a means of mitigation of these effects, that of sequential pixel display. 
    
    If we are using a sufficiently fast SLM, we can rely on the time averaging effects of the human eye to integrate multiple sub-frames. This is a technique that has been used extensively in previous holographic video research \cite{STTM}.
    
    \begin{figure}
        \centering
        {\includegraphics[trim={0 0 0 0},width=0.8\linewidth,page=1]{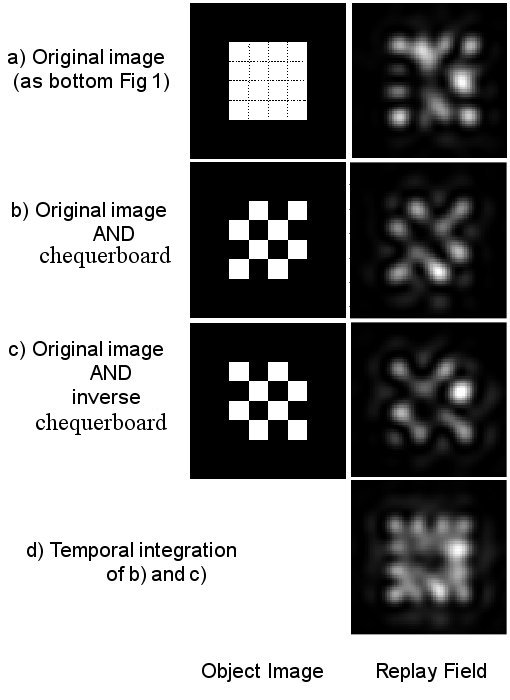}}
        \caption{Temporally alternating pixels to prevent coherent interference from neighbouring pixels and so improve image.}
        \label{fig:figure10}
    \end{figure}
    
    By separating the pixels so that horizontal neighbours are shown independently, we can reduce the amount of interdependence between neighbouring pixels. For example Figure~\ref{fig:figure10}a shows the case shown previously in Figure~\ref{fig:figure02}. This target image is then split into two portions in  (b) and (c), holograms generated for the portions and then the results shown in (d). Here it will be seen that the image quality observed is much higher.
    
    \begin{figure}
        \centering
        {\includegraphics[trim={0 0 0 0},width=0.8\linewidth,page=1]{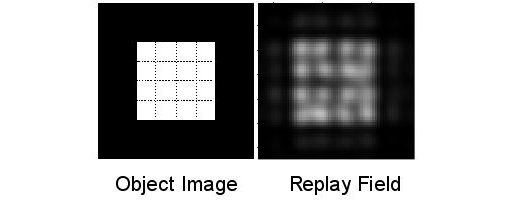}}
        \caption{Temporal integration of extra alternating frames so to prevent interference from diagonals as well.}
        \label{fig:figure11}
    \end{figure}
    
    This approach is taken further in Figure~\ref{fig:figure11} where diagonally neighbouring pixels are also shown independently leading to the sum of four independent sub-frames. Here the image quality is greatly improved with all pixels being clearly distinct at the expense of using 4 independent sub-frames.
    
    \subsection{Single Frame Manipulation}
    
    For applications where it is impractical to show sequential sub-frames, controlling phase relationships between pixels allows for improvements in pixel differentiation. This is in contrast to the random phase approach used earlier. As suggested in Figure~\ref{fig:figure03}, neighbouring pixels have the best independence with a pixel separation of either $\nicefrac{\pi}{2}$ or $\nicefrac{3\pi}{2}$.
    
    \begin{figure}
        \centering
        {\includegraphics[trim={0 0 0 0},width=0.8\linewidth,page=1]{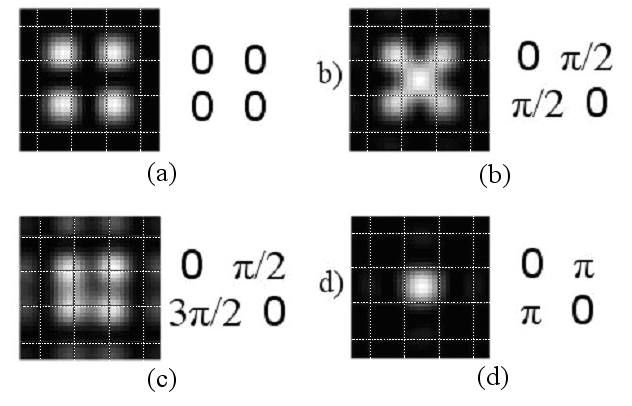}}
        \caption{Pixel differentiation for a $2 \times 2$ square of pixels with defined phase difference embedded in a continuous phase $128 \times 128$ pixel hologram. A single pixel grid is superimposed for reference.}
        \label{fig:figure12}
    \end{figure}
    
    Figure~\ref{fig:figure12} shows the case of a $2 \times 2$ block of neighbouring pixels with defined phase differences. The worst case of repelling is shown in (a) with the worst case of attraction in (d). The case in (b) shows the case of $\nicefrac{\pi}{2}$ neighbouring pixel separation but with similar diagonals and (c) shows a similar case but with diagonals separated by $\pi$. Here it can be seen that the defined phase of (c) leads to the best behaviour in pixel location but still involves significant error.
        
\begin{figure}
        \centering
        {\includegraphics[trim={0 0 0 0},width=0.8\linewidth,page=1]{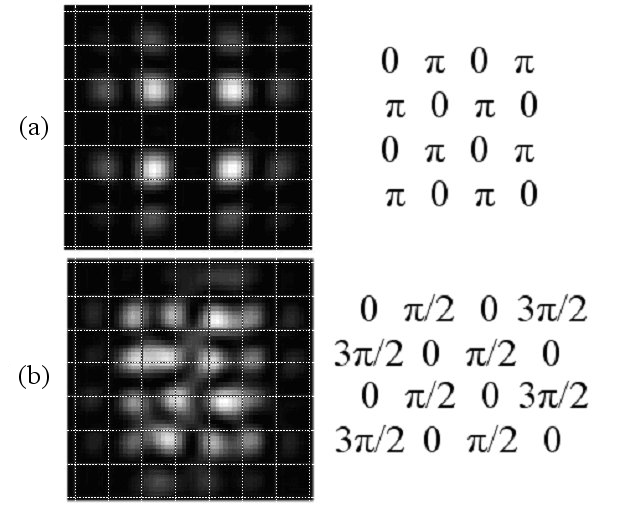}}
        \caption{Pixel differentiation for a $4 \times 4$ square of pixels with defined phase difference embedded in a continuous phase $128 \times 128$ pixel hologram. A single pixel grid is superimposed for reference.}
        \label{fig:figure13}
    \end{figure}

    We can extend this to the $4 \times 4$ block shown in Figure~\ref{fig:figure02} and with judicious tiling can get the behaviour shown in Figure~\ref{fig:figure13}. Again, while pixel differentiation is not perfect, it is significantly improved from Figure~\ref{fig:figure02}.
    
    \section{The effect of quantisation levels}
    
\begin{figure}
        \centering
        {\includegraphics[trim={0 0 0 0},width=0.8\linewidth,page=1]{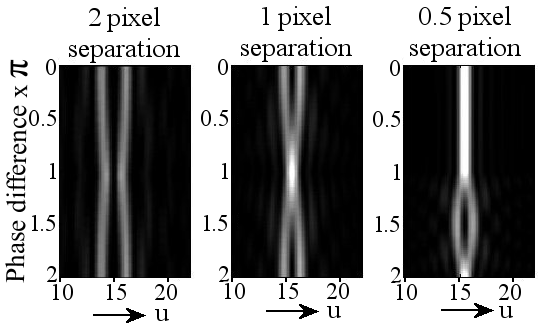}}
        \caption{Pixel differentiation for a $4 \times 4$ square of pixels with defined phase difference embedded in a continuous phase $128 \times 128$ pixel hologram. A single pixel grid is superimposed for reference.}
        \label{fig:figure15}
    \end{figure}

    The modeling shown until now has been using a continuous phase SLM. For real-world applications it is likely that these techniques will be used on multi-level or binary devices. If we ignore the effects of higher orders, we can reproduce the results of Figure~\ref{fig:figure03} for a binary phase SLM to give Figure~\ref{fig:figure15}. As before, the holograms are generated by superimposing constituent gratings and constraining the result to the SLM achievable states. Again, the pixels appear to move relative to each other.
    
    \section{Experimental Verification}
    
    \begin{figure*}
        \centering
        {\includegraphics[trim={0 0 0 0},width=0.6\linewidth,page=1]{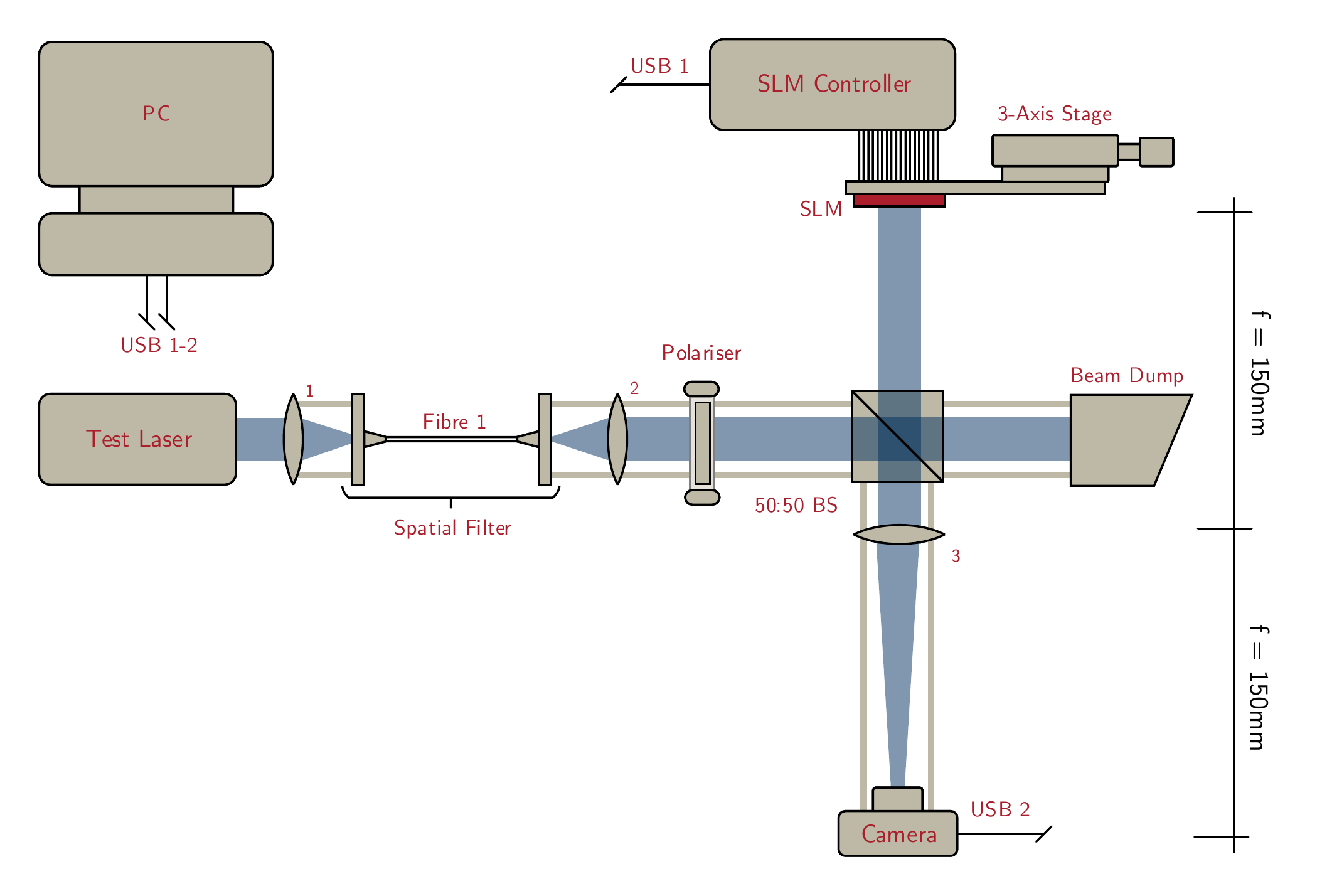}}
        \caption{Experimental configuration.}
        \label{fig:FreemanExperiment}
    \end{figure*}
    
    In order to confirm the modelling done, we showed the holograms generated for Figure~\ref{fig:figure15} on a binary Electrically Addressable Spatial Light Modulator (EASLM) in a configuration shown in Figure~\ref{fig:FreemanExperiment}. The EASLM available was a CRLOpto (now Forth Dimension) binary device with $1280 \times 1024$ pixels and $13.6 \mu{}m$ pixel pitch. We showed the binary holograms on the central $128 \times 128$ pixel region and and confirmed flatness using a Fabry-Perot  interferometer. The hologram was displayed using a 2f system with a 150mm focal length doublet focussing the replay image onto a CCD camera. The row of pixels from the camera, passing through the replay image of both dots was then extracted and used to build the plot.
    
    \begin{figure}
        \centering
        {\includegraphics[trim={0 0 0 0},width=0.8\linewidth,page=1]{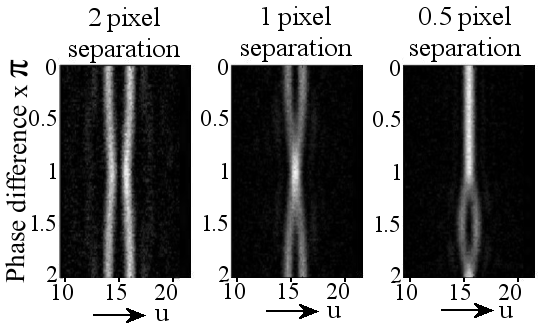}}
        \caption{Experimental confirmation of pixel differentiation results.}
        \label{fig:figure18}
    \end{figure}

    It will be noticed that there is a high degree of similarity between the theoretical model and the experimental results with the addition of some additional noise.
    
    \section{Summary and Recommendations}
    
    This paper has presented the observation of the interplay between neighbouring pixels in CGH. Whereas the replay field sampling is equivalent to the DFT of a given SLM aperture function, the interstitial behaviour can vary significantly. By summing the individual SLM pixel contributions for interstitial points we can get a higher resolution understanding of individual replay field pixels.
    
    Our main observation in this work is that neighbouring pixels appear to repel or attract each other depending on their relative phase differences. We have presented a mathematical argument that this is be due to a form of beating behaviour between the individual pixel contributions. 
    
    This effect was seen to be independent of the location of the pixel and was only influenced by their relative phase difference and their relative spacing. Additionally, the effect was independent of the image resolution.
    
    In the light of this we suggested two mitigation techniques. The first involved time-averaging of neighbouring pixels. Doing this allowed for much clearer pixel differentiation at the cost of requiring additional sub-frames. The second involved replacing the phase randomisation step with a defined phase difference between neighbouring pixels. 
    
    We then extended our results to the case of a binary phase SLM demonstrating similar results before introducing an experimental system. This confirmed the pixel movement observed.
    
    Often when comparing an experimental holographic image with a theoretically expected result the observed noise is assumed to be attributable to a variety of sources including: speckle, non-flatness of the SLM, beam shape and atmospheric effects. On the basis of this work, we suggest that even in a theoretically perfect system, the sampled replay field can be a poor indicator of the actual image displayed. 
    
    The effects we have shown here of pixel movement is not observable in a sampled replay field and that interstitial behaviour in the replay field should be considered when discussing algorithms. It is suggested that some of the experimental noise observed in real-world systems may be attributed to hologram generation approach used rather than imperfections in the physical system. 
    
    While we have discussed here a simple example of apparent pixel movement, we are aware that many other complex effects are likely and further research is recommended.
         
    \section{Conclusion}
    
    In this work we have presented a theoretical investigation of interstitial noise in hologram generation and have highlighted one particular area. That of pixel differentiation. We have examined this phenomena in detail discussing variables that influence behaviour. Informed by this we have presented two techniques for mitigation and given recommendations for future study.
    
    \subsection*{Acknowledgements}
    
    This paper is based on an unpublished work "Resolution in the Replay Field of a Holographically Generated Video Image" by JPF and TDW. PJC and RM are responsible for updating and extending this paper and have done so with permission.
    
    \subsection*{Funding}
    
    PJC and RM would like to thank the Engineering and Physical Sciences Research Council (EP/L016567/1 and EP/L015455/1) for funding through the period of this research.
    
    JPF would like to thank BAE Systems for sponsorship and support.
    
    \subsection*{Disclosures}
    
    The authors declare no conflicts of interest.
    
    \bibliography{references}
    \bibliographystyle{spiejour} 
    
\end{document}